\documentclass[journal]{IEEEtran}

\usepackage{multirow}
\usepackage{graphicx}
\usepackage[table,xcdraw]{xcolor}
\usepackage[cmex10]{amsmath}
\usepackage{cite}
\usepackage{amsmath,amssymb,amsfonts}
\usepackage{siunitx}    % v3!
\usepackage{ragged2e}
\usepackage{caption} 
\usepackage{booktabs}
\usepackage{hyperref}
\usepackage{booktabs}
\usepackage{mwe}
\usepackage{booktabs, makecell, multirow, tabularx}
\setcellgapes{2pt}
\newcolumntype{L}[1]{>{\RaggedRight\hspace{0pt}%
                     \hsize=#1\hsize}X}
\usepackage{lipsum}
\usepackage{atbegshi}% http://ctan.org/pkg/atbegshi
\AtBeginDocument{\AtBeginShipoutNext{\AtBeginShipoutDiscard}}
\begin{document}
\bstctlcite{IEEEexample:BSTcontrol}
    \title{Multi-channel polarization manipulation based on graphene for encryption communication}
  \author{Parsa~Farzin,
      Mohammad Javad Hajiahmadi,
      and Mohammad~Soleimani*}

  \thanks{School of Electrical Engineering, Iran University of Science and Technology, Tehran, 1684613114, Iran}


% ====================================================================
\maketitle

% === ABSTRACT ====================================================================
% =================================================================================
\begin{abstract}
%\boldmath
Wave-based cryptography, at the vanguard of advancing technologies in advanced information science, is essential for establishing a diverse array of secure cryptographic platforms. The realization of these platforms hinges on the intelligent application of multiplexing techniques, seamlessly combined with appropriate metasurface technology. Nevertheless, existing multi-channel encryption technologies based on metasurfaces face challenges related to information leakage during partial channel decoding processes. In this paper, we present a reprogrammable metasurface for polarization modulation. This metasurface not only allows for the arbitrary customization of linearly polarized reflected waves but also enables real-time amplitude modulation. Here, relying on polarization amplitude control, a fully secure communication protocol is developed precisely in the terahertz (THZ) spectrum to achieve real-time information encryption based on polarization modulation metasurfaces where access to information is highly restricted. The proposed metasurface employs the double random phase encryption (DRPE) algorithm for information encryption. It transmits the encrypted data through different polarization channels using two graphene nanoribbons, exclusively controlled by external biasing conditions. Various encryption scenarios have been outlined to fortify information protection against potential eavesdroppers. The simulated results show that this unique technology for hiding images by manipulating the polarization of the reflected wave provides new opportunities for various applications, including encryption, THz communications, THz secure data storage, and imaging.
\end{abstract}

% === KEYWORDS ====================================================================
% =================================================================================


% \fi
\IEEEpeerreviewmaketitle

% ====================================================================
% ====================================================================
% ====================================================================

% === I. INTRODUCTION =============================================================
% =================================================================================
\section{Introduction}

\IEEEPARstart{T}{erahertz} (THz) radiation denotes electromagnetic (EM) waves with frequencies spanning from 0.1 to 10 THz, positioning them in the spectrum between infrared waves and microwaves \cite{1}. In the past decades, THz wave manipulation has been of great importance for creating a new generation of information technologies due to its unique advantages in terms of high bandwidth and high resolution, which has attracted great attention in wireless communication, remote sensing, and holographic communication \cite{2}. Nevertheless, the interaction between THz radiation and most natural materials has revealed weak EM responses, leading to a lack of technological advancements and practical applications in THz devices \cite{3}.

Metasurfaces, which are two-dimensional (2D) metamaterials consisting of engineered subwavelength periodic or nonperiodic geometric structures, have been intensively investigated also to offer advantages such as  their unique properties in manipulating EM waves, cost-effectiveness, high surface integrity, and simplified fabrication processes. \cite{4,5,6}. As an emerging platform, metasurfaces have sparked a surge in research interest by showcasing exceptional capabilities in manipulating EM wave amplitude, phase, and polarization. These advancements hold significant promise for diverse applications in EM wave control, including the generation of orbital angular momentum (OAM) \cite{7}, polarization conversion \cite{8}, and holography \cite{9}. In a groundbreaking development, T. Cui et al. have introduced the revolutionary concept of digital metasurfaces, serving as a bridge between the physical and digital realms. This innovation opens up the opportunity to reconsider metasurfaces from the standpoint of information science \cite{10}. In the digital coding paradigm, Typical physical parameters such as amplitude, phase, and polarization can be represented as digital states "0" or "1" \cite{11,12,13}. Leveraging a field-programmable gate array (FPGA), these coding metasurfaces can be digitized and controlled programmatically in a real-time communication \cite{14}.

OOn the other hand, manipulation of polarization , a fundamental aspect of EM waves, is also of great significance for practical applications, ranging from communication to imaging \cite{15,16,17}. Polarization Modulation, employing the polarization mode of EM waves as an information-bearing parameter, has been suggested as a viable alternative to conventional modulation techniques, including amplitude-shift keying, phase-shift keying, and frequency-shift keying \cite{18}. Several versatile methods for Polarization Modulation, relying on passive metasurfaces and employing the principles of phase control—such as geometric phase\cite{19} and propagation phase\cite{20}—have been proposed. However, a limitation lies in the fact that the functions of these passive metasurfaces remain fixed once manufactured. Fortunately, by using active components (such as graphene \cite{21,22}, vanadium dioxide ($VO_2$) \cite{23}, and liquid crystals \cite{24}) tunable metasurfaces can be achieved to realize dynamical control of EM waves. It is important to highlight that these tunable metasurfaces have been utilized to dynamically manipulate the real-time polarization state of EM waves or light \cite{25,26,27}. Hence, in wireless communication, the utilization of these metasurfaces allows for the transmission of encrypted information through various polarizations.

\begin{figure*}
    \centering
    \includegraphics[scale=.4]{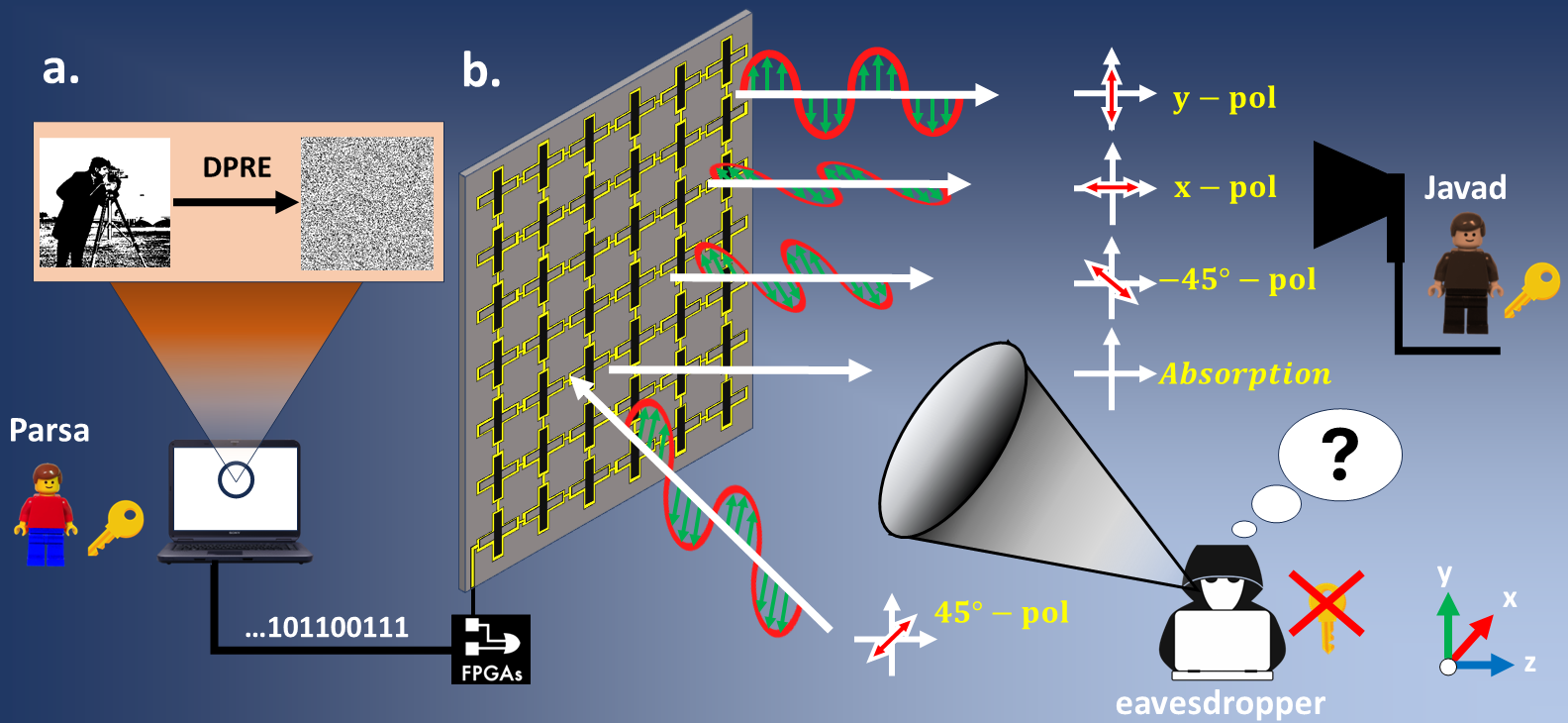}
    \caption{Conceptual illustration of the programmable polarization-multiplexed information metasurface. (a) The target image, the cameraman, is encrypted using the algorithm of Double Random Phase Encryption (DRPE) and the binary bits are recalled through a field programmable gate array (FPGA) for the sequential configuration of the proposed metasurface. (b) The metasurface, which employs voltage-controlled polarization modulation, conveys information via distinct polarization channels. The information received is then decryption on the receiving end.}
    \label{fig1}
\end{figure*}

Over the years, ensuring information security has consistently been crucial in communications. Particularly in the digital age of contemporary society, safeguarding the privacy of information is essential for alleviating concerns related to data sharing and the potential misuse of important data \cite{28}. Recognizing the significant importance of information security, numerous encryption techniques have been proposed and developed \cite{29,30,31,32}. However, traditional encryption techniques depend on nonlinear optics, which entails high intensity and energy requirements. Modern communication is evolving towards wave-based cryptography, showcasing significant potential in enhancing information security. This progression is attainable through the abundant degrees of freedom linked to various optical channels, such as frequency \cite{33}, amplitude \cite{34}, phase \cite{35}, and polarization \cite{36}. To increase the information capacity and encryption security in metasurfaces, intelligent multiplexing techniques (such as propagation direction multiplexing \cite{37}, wavelength multiplexing \cite{38}, angular multiplexing \cite{39} and polarization multiplexing \cite{40}) are in great demand. In this scenario, encrypted image (or images) or text information is transmitted through diverse channels. Among these approaches, polarization multiplexing is a highly suitable method for transmitting encrypted information through different polarizations. In previous studies, information encryption through different polarization channels has been explored \cite{41,42,43,44,45}. However, the predominant emphasis has been on static metasurface devices, where the properties of individual pixels cannot be reconfigured in real-time, especially in the THz spectrum \cite{46}. Hence, the endeavor to achieve a re-configurable polarization-multiplexed metasurface remains a formidable challenge, and this objective is still far from being well addressed or solved. Overcoming the challenges in real-time meta-cryptography involves tasks such as designing a polarization-multiplexed metasurface with subwavelength pixels and achieving independent control of dynamic pixels simultaneously.

In this paper, we propose a reprogrammable metasurface for polarization modulation capable of not only achieving arbitrary linearly polarized reflected waves but also dynamically modulating the amplitude of the reflected waves in real-time. Furthermore, the polarization modulation information metasurface is proposed multiple communication encryption scenarios, where the transmitted information is encrypted using the double random phase encryption (DRPE) algorithm. The proposed meta-atom is composed of dual layers of gold Jerusalem cross slits and two layers of graphene nanoribbons. Through the dynamic tuning of the Fermi energy of each graphene bilayer, real-time control over the amplitude of both linear polarizations is achieved. The transmission of encrypted information occurs via dynamically modulated polarization channels, achieved by tuning the Fermi energy of the graphene layers a through biasing cotroller (FPGA). For the transmission of encrypted information across distinct polarization channels, we manipulate two domain-level states for both x- and y-polarizations, resulting in a 2-bit meta-atom. The results demonstrate that the proposed rewritable information metasurface not only attains excellent polarization modulation performance but also paves the way for enhancing information security with a high degree of flexibility in multi-channel information encryption, THz data storage, THz communication, and information processing.

\section{Reprogrammable Polarization Modulation Based on Information Metasurface} 

The majority of existing design methodologies depend on full-wave numerical simulations to correlate a specific sub-wavelength element with its corresponding digital status. This process results in the creation of a complex lookup table, which serves as the foundation for coding metasurface realization \cite{47}. On the contrary, the primary objective of this study is to design a coded metasurface where the operational state of full structure can be individually configured through a DC voltage source, providing dynamic and real-time control over the polarization of the reflected wave at THz frequencies. Figure 1 depicts a conceptual illustration of the communication encryption scheme utilizing the programmable polarization-multiplexing information metasurface. This illustration presents a schematic of the proposed re-writable metasurface, comprising digital lattices based on graphene, designed for polarization modulation of the incident wave. It's noteworthy that, during the transmission of information, a target image, like a photographer's image, is encrypted by using DRPE algorithm and is sent from Parsa with different polarizations, as shown in Figure 1a. The proposed metasurface consists of two perpendicular nanoribbon graphenes, which can independently control the amplitude of vertical and horizontal polarizations in real time by adjusting the state of the chemical potential of graphene, as shown in Figure 1b. Hence, when a 45°-polarized incident wave illuminates to the metasurface, the polarization state of the linearly polarized reflected EM wave can be freely modified. The ability to dynamically control the polarization of the reflected wave enables us to transmit encrypted information through various polarization channels using DRPE, significantly enhancing the security of the channel. At the receiving end, javad, using the keys that are only available to him and Parsa, decrypted the encrypted target information. Hence, even if a snooper intercepts the encrypted information, deciphering it without the key is an impossible task. 

\section{Design of the Reprogrammable Polarization Information Metasurface} 
\subsection{Complex Surface Conductivity of Graphene}

\begin{figure*}
    \centering
    \includegraphics[scale=.3]{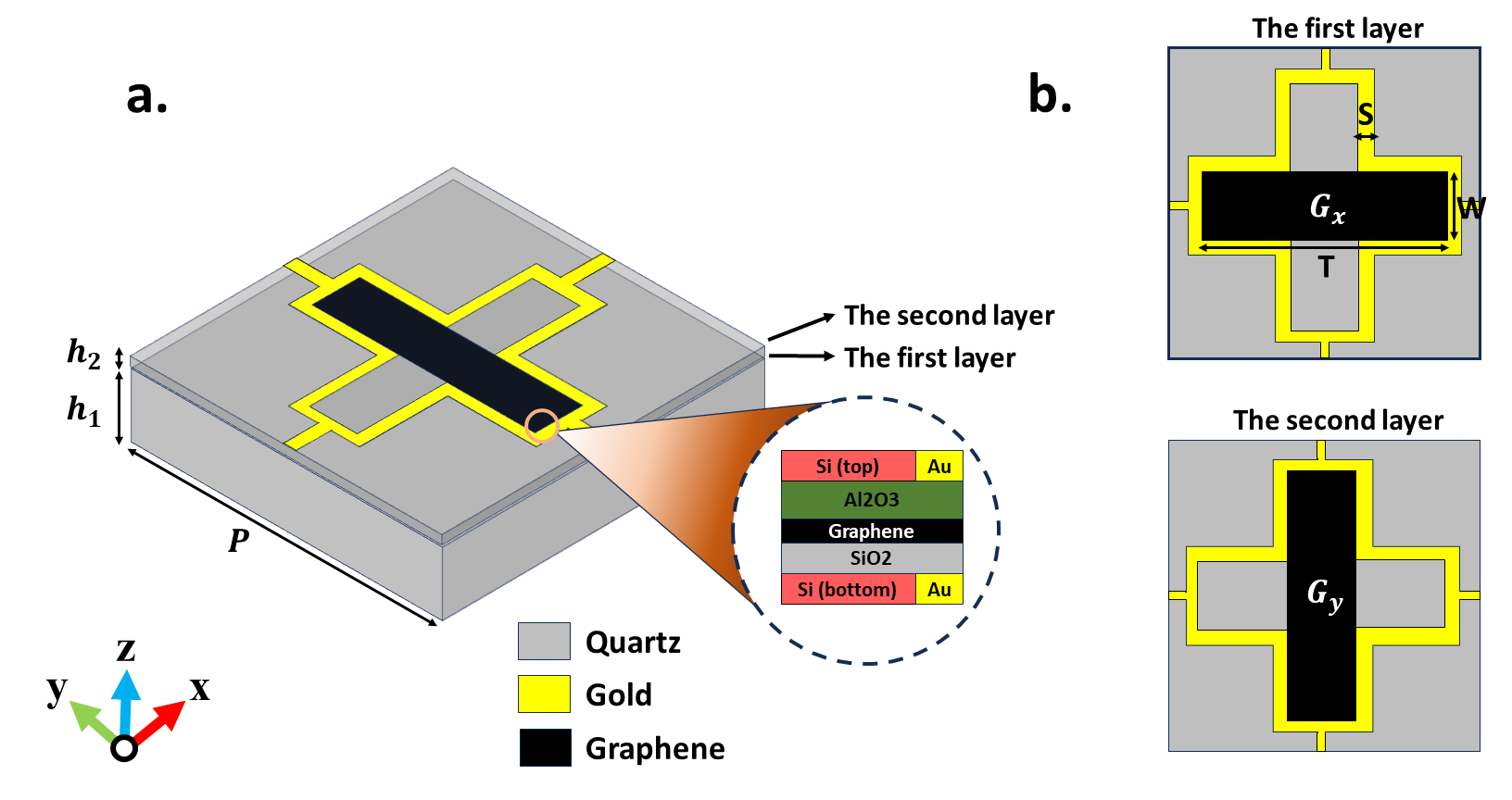}
    \caption{Reprogrammable meta-atom geometry. a) The proposed meta-atom comprises two layers of graphene, with each layer featuring a Si/SiO2/Graphene/Al2O3/Si layered structure. b) Front view of both layers.}
    \label{fig2}
\end{figure*}

Graphene comprises a single layer of carbon atoms organized in a honeycomb lattice structure. Being a zero-gap semiconductor with distinctive electrical, thermal, optical, and mechanical properties, graphene has garnered considerable attention over the past decade, especially in the THz band \cite{48}. Furthermore, owing to its monoatomic structure, graphene can be locally represented by a complex surface conductivity tensor \cite{49}:

\begin{equation}
   \sigma(\omega, \mu_c(E_0 ), \Gamma, T, B_0 )= \hat{x}\hat{x}\sigma_{xx} + \hat{x}\hat{y}\sigma_{xy} + \hat{y}\hat{x}\sigma_{yx} + \hat{y}\hat{y}\sigma_{yy}
   \label{eq1}
\end{equation}

where $\omega$ is the radian frequency, $\Gamma=1⁄2\tau$ is the phenomenological scattering rate with $\tau$ being the electron-phonon relaxation time. T is the room temperature, $\mu_c$ is the chemical potential, $E_0$ and $B_0$ are electrostatic and magnetostatic bias fields, respectively. Without magnetostatic bias, the off-diagonal components of the surface conductivity tensor become vanished, leading to the observation of isotropic behaviors in the graphene monolayer. In accordance with the Kubo formula, the complex surface conductivity of graphene can be expressed by considering both interband and intraband transitions \cite{50}:

\begin{equation}
\sigma_s (\omega, \mu_c, \Gamma, T) = \sigma^{\text{intra}}_s (\omega, \mu_c, \Gamma, T) + \sigma^{\text{inter}}_s (\omega, \mu_c, \Gamma, T)
   \label{eq2}
\end{equation}

\begin{equation}
    \sigma^{\text{intra}}_s (\omega, \mu_c, \Gamma, T)=-\frac{ie^2 k_B T}{\pi \hbar^2 (\omega-i/\tau)}
    [\frac{\mu_c}{k_B T} + 2ln(e^\frac{-\mu_c}{k_B T}+1)]
    \label{eq3}
\end{equation}

\begin{equation}
    \sigma^{\text{inter}}_s (\omega, \mu_c, \Gamma, T) = -\frac{ie^2}{4\pi \hbar} ln[\frac{2|\mu_c |-(\omega-i⁄\tau)\hbar}{2|\mu_c |+(\omega-i⁄\tau)\hbar}]
   \label{eq4}
\end{equation}

Here, $\hbar=h/2\pi$ represents the reduced Plack’s constant, $e=1.6\times10^{-19}$C is electron charge, $k_B$ is the Boltzmann’s constant, and $\mu_c$  is the chemical potential. 
In the low THz frequency region, the interband contribution to graphene conductivity can be disregarded due to the Pauli exclusion principle, given that the photon energy is much less than the Fermi energy $\hbar \omega \ll E_f$ \cite{51}. In this paper, we held temperature and relaxation time constant at T = 300 K and $\tau$ = 1 ps, respectively, throughout the entirety of this study. It is worth mentioning that the surface conductivity of the graphene surface can be dynamically altered by an external electrical bias (Further details can be found in Supplementary A), enabling real-time implementation of all the different functions for the proposed metasurface mentioned in Figure 1.

\subsection{Meta-Atom Design Principle and Simulation Results}

Figure 2(a) depicts the suggested meta-atom designed for polarization modulation, comprising two layers of gold with floating gates based on two graphene nanoribbons, all sandwiched between two layers of quartz. Floating gates are employed to adjust the charge density of each graphene layer by varying the DC bias voltage, encompassing $Si$, $SiO_2$, $Al_2O_3$, and graphene nanoribbons (detailed information can be found in Supplementary Information B). The substrates within the floating gate, owing to their significantly small thickness compared to the operating wavelength, can be disregarded for their negligible impact on the amplitude and phase of the reflected waves. In each layer a gold Jerusalem cross slot exists with the graphene nanoribbons positioned horizontally in the first layer ($G_x$) and vertically in the second layer inside the Jerusalem cross slot ($G_y$), as shown in Fig 2(b). The key parameters for the design of the Jerusalem cross slot and graphene nanoribbons are S=1 µm, T = 20 µm, and W=6 µm, respectively. The quartz dielectric utilized in this structure exhibit a relative permittivity ($\epsilon_r$) of 3.75 and a loss tangent (tand) of 0.0004, with thicknesses of $h_1$ = 33 µm and $h_2$ = 33.4 µm , respectively. Each meta-atom possesses a periodicity of P=50 µm and is terminated by a gold plane to impede the transmission of energy into the rear of the Information Metasurface. Due to the connection of the Jerusalem cross slot of each of the meta-atoms to each other through gold, the proposed metasurface biasing will be very convenient, and only by biasing one of the meta-atoms, all the meta-atoms formed in the metasurface can be biased (for more information, see Supplementary information can be found in Figure S1). The simulations for each meta-atom were carried out using CST Microwave Studio, a sophisticated full-wave commercial software. Periodic boundary conditions were employed in both the x- and y-directions, and Floquet ports were additionally assigned in the z-direction.

The polarization of the reflected wave from each individual element can be independently controlled by adjusting the chemical potential of each graphene ribbon. Figure 3 shows the simulation results of the control of reflected wave amplitude for different chemical potentials under waves with 45° polarization incidences. Utilizing chemical potential on graphene ribbons to accomplish various functions is illustrated as $\mu_c = \{\mu_c^{G_x},\mu_c^{G_y}\}$, where $\mu_c^{G_x}$ denotes the chemical potential applied to $G_x$, and $\mu_c^{G_y}$ represents the chemical potential applied to $G_y$. As depicted in Figures 3a to 3d, to attain absorption for both polarizations, x-, y-, and -45°-polarization at the frequency of 2.04 THz, the chemical potential values for each $G_x$ and $G_y$ are represented as [\{0.345 eV, 0.345 eV\}, \{0 eV, 1 eV\}, \{1 eV, 0 eV\}, \{0 eV, 0 eV\}], respectively. It's important to highlight that in the -45° polarization reflection mode, the amplitude and phase are identical. Consequently, it becomes feasible to independently and in real-time control the amplitude as 0 and 1 for each polarization, achieving a 2-bit polarization state. In this work, absorption for both polarizations, x-, y-, and -45°-polarization channels, are employed in the encryption scheme, which are encoded as polarization codes of “00”, “01”, “10”, and “11”, respectively. The results and explanations concerning the proposed metasurface, comprised of 20 × 20 elements, are provided in supplementary section C. 

The meta-atom that has been designed can be depicted through a physically equivalent circuit (see Supplementary Information D). In this representation, each structured graphene pattern is encircled by a Jerusalem cross slit, illustrated as an R-L-C branch. Utilizing the circuit model, we computed the amplitude modulation for both x- and y-polarizations. The determined values for R, L, and C are provided in Supplementary Table S1 for different scenarios. Supplementary Fig. S5(a)–(h) compares the spectra derived from the circuit model with those from full-wave numerical simulations, demonstrating a excellent agreement in amplitude modulation for both x- and y-polarizations. As a result, the development of the circuit model enhances the proposed metasurface design process, and the achieved results closely match our expectations.

\begin{figure*}
    \centering
    \includegraphics[scale=.4]{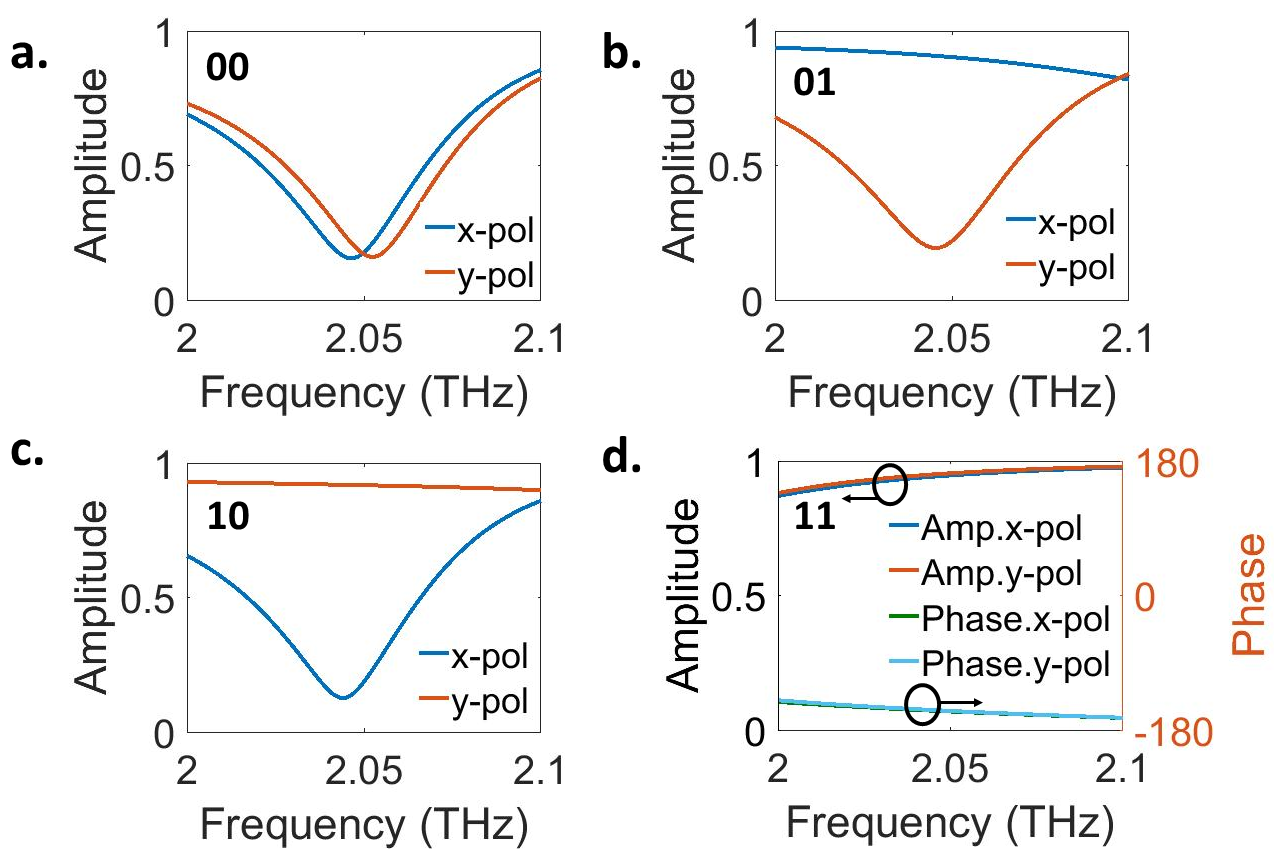}
    \caption{Simulated amplitude modulation for each of x- and y-polarizations. (a) Absorption mode, (b) reflection mode with x-polarization, (c) y-polarization, (d) -45°-polarization.}
    \label{fig3}
\end{figure*}

\section{Implementation of metasurface-assisted polarization modulation for advanced information encryption}

Optical encryption serves as a safeguard against attacks on sensitive information. Among the optical encryption methods, DRPE stands out for its ability to deliver swift encryption and decryption, coupled with a straightforward implementation process. The proposed metasurface is designed to establish independent polarization channels housing seemingly undiscoverable images. This significantly enhances the security level of both information encryption and data storage. Utilizing a robust encryption algorithm known as $cipher$ \cite{52}, various combinations of letters (plain-texts images) are encrypted into a series of hotspots (ciphertext image) within each channel. The cipher image undergoes conversion into reflection polarization profiles through the DPRE algorithm, subsequently being encoded into the biasing voltages applied to the vertical graphene on the metasurface. Also, Double-phase random encryption is an advanced cryptographic technique designed to enhance the security of sensitive data. Unlike traditional encryption methods, which rely on a single layer of protection, DRPE introduces an additional layer, significantly bolstering the overall security posture (further details are available in supplementary material section E). Hence, utilizing DRPE and 2-bit polarization control enables us to transmit diverse information, including images with high security and various scenarios.

Let us consider the following scenario. In the first scenario, illustrated in Figure 4, Parsa aims to securely and privately transmit a message containing a 1-bit image, richard feynman, to Javad without compromising integrity. In this instance, utilizing the DRPE algorithm, he transforms the plaintext image into a binary ciphertext image featuring white and black dots. To enhance the security of sending the desired information, Parsa transforms the ciphertext image into three separate ciphertext images, and a mathematical relationship exists between these three ciphertext images and the original ciphertext image  (illustrated by three ciphertext images labeled as ciphertext images 1, ciphertext images 2, and ciphertext images 3). The mathematical relationship is such that each pixel in the original ciphertext image equals the XOR of the corresponding pixels in the three ciphertext images (original ciphertext image(i,j) = ciphertext image 1(i,j) $\oplus$ ciphertext image 2(i,j) $\oplus$ ciphertext image 3(i,j), where $\oplus$ is the XOR symbol, and (i,j) is the coordinates of the pixels of the ciphertext images). Subsequently, each of the three ciphertext images is assigned to a specific polarization channel. Ciphertext image 1 is allocated to x-polarization, ciphertext image 2 to y-polarization, and ciphertext image 3 to -45°-polarization. In this case, we define the white dots to represent bit "1" and the black dots to represent bit "0". Each of the three cipher text images is converted into a sequence of binary symbols, represented by "0" and "1". Subsequently, each of the three ciphertext images is assigned to a specific polarization channel. 

\begin{figure*}
    \centering
    \includegraphics[scale=.18]{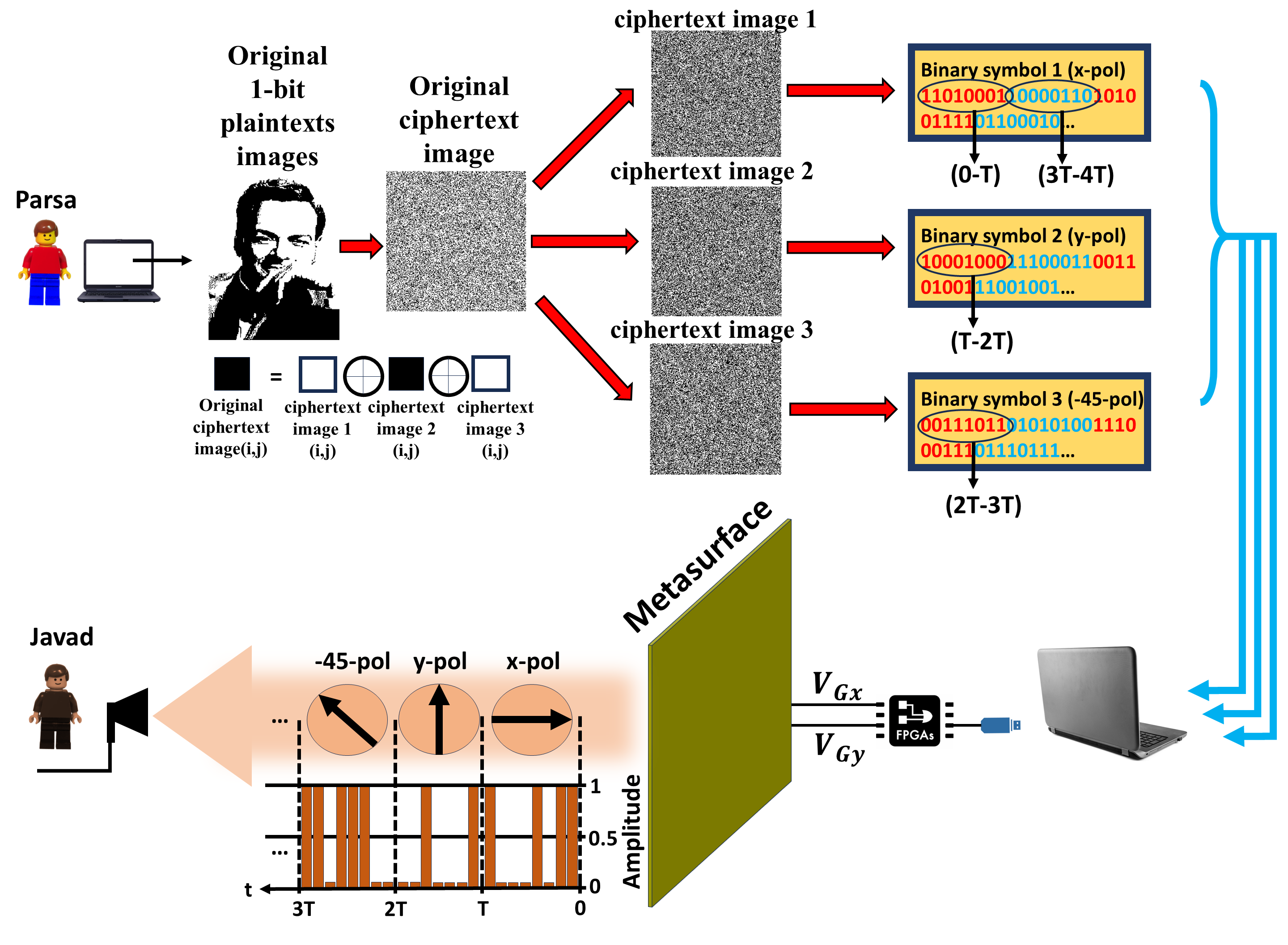}
    \caption{The process of encrypting a 1-bit image and transmitting it through various polarization channels.}
    \label{fig4}
\end{figure*}

Ciphertext image 1 is allocated to x-polarization, ciphertext image 2 to y-polarization, and ciphertext image 3 to -45°-polarization. Consequently, we reassign the meanings of the codes "0" and "1," where "0" indicates a low-level signal, and "1" represents a high-level signal. These codes correlate with the amplitude of the EM wave in each polarization channel, where code "0" is associated with low reflection, and code "1" with high reflection. It is important to note that the definition of bit "1" varies across all three ciphertext images. Specifically, bit "1" represents the high amplitude of x-polarization in ciphertext image 1, y-polarization in ciphertext image 2, and -45°-polarization in ciphertext image 3. It's crucial to emphasize that in all three ciphertext images, the "0" bit denotes wave absorption, indicating the absence of transmitted information during such instances. Sending the binary information obtained from encrypting Richard Feynman's image from the computer to the FPGA and adjusting the chemical potential of graphene through it, enables the immediate adjustment of the reflected wave's polarization for the transmission of information in encrypted form. For the transmission of encrypted data, we divide each of the three sequences of binary symbols into 8-bit segments, denoted in red and blue. In this scenario, the initial 8 bits of each of the three images correspond to the x-, y-, and -45°-polarizations during the time intervals 0-T, T-2T, and 2T-3T, respectively. Subsequent 8-bit segments are transmitted in the succeeding time intervals. On the other hand, Javad, which has a receiver that is sensitive to both x- and y-polarizations, receives different polarizations at different time intervals. Because he possesses two decryption keys exclusive to himself and Parsa, he decrypts the information received from Parsa and reverses any data encryption on the sender's side on the receiver's side (detailed information about decryption can be found in Supplementary Information Figure S6). Ultimately, upon decrypting the received data, Javad recognizes that the transmitted image is from Richard Feynman. In this scenario, the transmission of information becomes intricate with the presence of three ciphertext image and the utilization of various polarization channels. Breaking this encrypted information would pose a considerable challenge for any potential eavesdropper.

In the first scenario, Parsa had the capability to transmit just one encrypted image to Javad by leveraging the polarization modulating metasurface. Therefore, one of the upcoming challenges is to send more than one photo encrypted so that the eavesdropper does not notice the data sent between the channels. In the second scenario, two 1-bit images are merged, and the information are encrypted. As illustrated in Figure 5, Parsa intends to send two 1-bit images (depicting a tiger and a male cameraman) via distinct polarization channels. When merging two images, if we treat two 1-bit images as two raster planes of identical dimensions, each pixel carries 1 bit of information (either bit 0 or 1, where bit 0 signifies black points and bit 1 signifies white points). By superimposing these two planes, each pixel undergoes a transformation from 1-bit to 2-bit (0+0=00, 0+1=01, 1+0=10, 1+1=11). In this case, the image will no longer exhibit black and white dots. With the transition to a 2-bit image, the colors now encompass black ("00"), deep gray ("01"), pale gray ("10"), and white ("11"). Consequently, this process leads to the conversion of two 1-bit images into a single 2-bit image through their amalgamation. Subsequently, we encrypt the acquired 2-bit image using DRPE. In contrast to the first scenario, the resulting encrypted image is a 2-bit representation that encompasses the colors of the combined image. We convert the bits obtained from the sum of two pictures into a sequence of binary symbols. Presently, this encrypted information, akin to the first scenario, is transmitted to an FPGA. Subsequently, by adjusting the chemical potential of graphene through the FPGA, various and desired polarizations can be manipulated to facilitate the transfer of information in coded form. We consider the bits assigned to the combined image with the bits assigned to the control results of the modulation amplitude (Figure 2) proportional to each other (absorption=00, x-pol=01, y-pol=10, -45°-pol= 11). Therefore, to transmit encrypted information, we partition the binary symbol sequence into two segments and send each set of bits within its respective amplitude. On the receiver side, Javad, similar to the first scenario, receives the encrypted data through different polarization amplitudes. He receives different polarizations successively and receives the encrypted data. Javad decrypts the received information using a shared key exclusively between him and Parsa, performing the inverse merge of the two images (detailed information about decryption can be found in Supplementary Information Figure S6). Ultimately, upon decryption the received data, Javad discerns that the transmitted information comprises images of both a tiger and a male photographer. Up to this point, the scenario of encrypting one or two images and transmitting them through different polarization channels are examined. A comprehensive discussion on alternative scenario for transmitting more than two images is available in the supplementary Information F.

\begin{figure*}
    \centering
    \includegraphics[scale=.18]{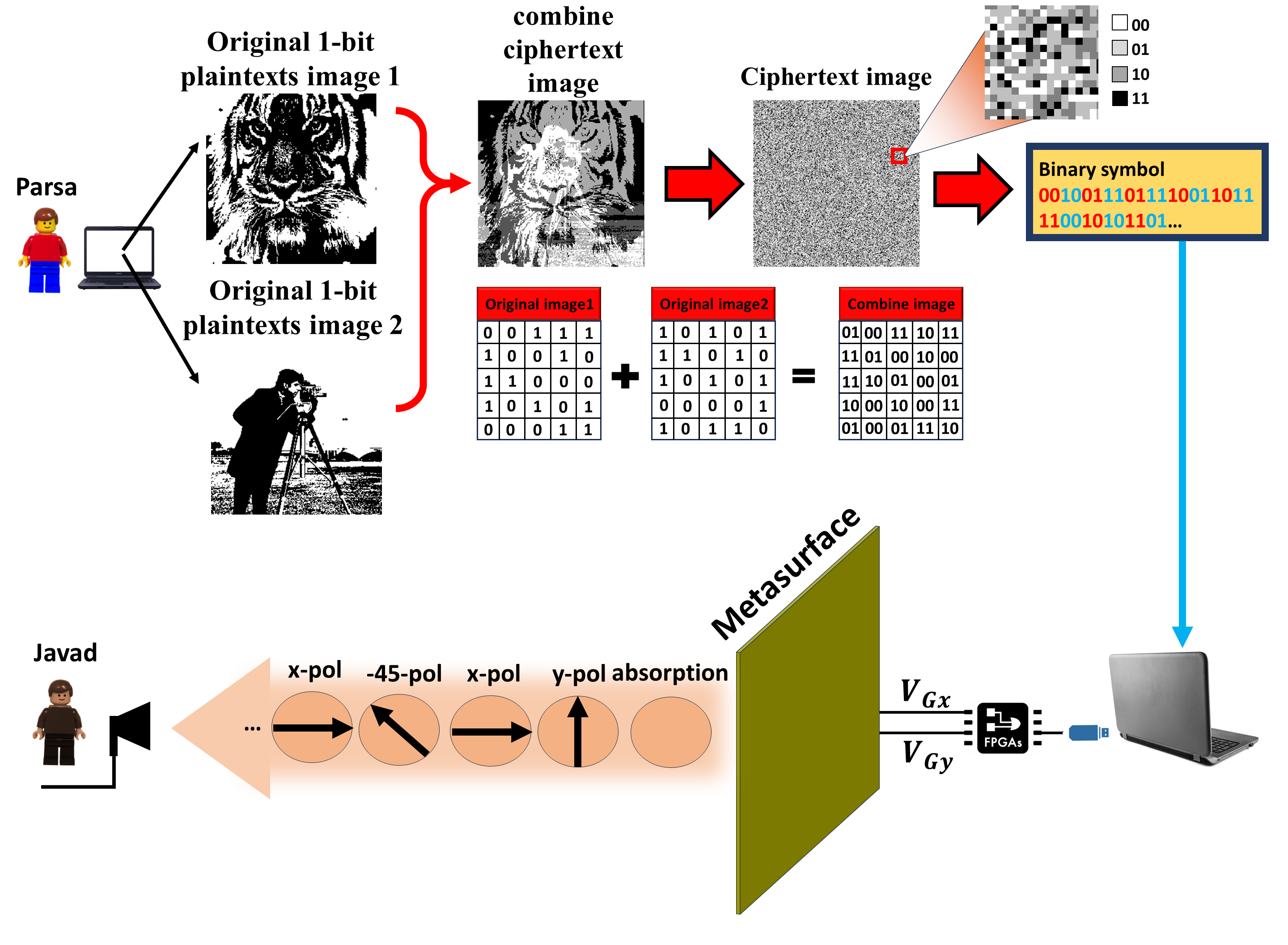}
    \caption{The procedure of encrypting two 1-bit images and transmitting them through amplitude modulation in both the x- and y-polarization channels.}
    \label{fig5}
\end{figure*}

\section{Conclusions}
In this paper, we presented a reprogrammable metasurface for polarization modulation. This metasurface not only achieves the desired polarized reflected waves but also dynamically controls the amplitude of the reflected waves. Employing the polarization modulation information metasurface, we encrypted one or two images in two distinct scenarios using the DRPE algorithm. Subsequently, these encoded images were transmitted through varying polarization channels and amplitudes. The suggested meta-atom comprises dual layers of Jerusalem gold cross slits and dual layers of graphene nanoribbons. By applying an external electric bias to each graphene layer, real-time control over the amplitude of both linear polarizations could be achieved using a FPGA. The information encrypted by DRPE is transmitted through four distinct channels, each featuring two domain-level modes in both linear x- and y-polarizations. Ultimately, on the receiver side, the received information undergoes decoding through the reverse process DRPE and the use of the corresponding keys. The proposed metasurface, with its reprogrammability, information encryption capabilities, and multi-channel nature, holds significant promise in the fields of multi-channel information encryption, THz communication, and information processing.

% or


% ============================================
%\appendices
%Appendix one text goes here %\cite{Roberg2010}.

%\section{}

%\section*{Acknowledgment}

%The authors would like to thank D. Root for the loan of the SWAP. The SWAP that can ONLY be usefull in Boulder...

\ifCLASSOPTIONcaptionsoff
  \newpage
\fi



%\begin{thebibliography}{1}


\bibliographystyle{IEEEtran}
\bibliography{IEEEabrv,MTT_reveyrand}

\vfill


\end{document}